# Light-induced quantum anomalous Hall effect on the 2D surfaces of 3D topological insulators

Haowei Xu [1], Jian Zhou [1], and Ju Li [1,2] [†]

[1] Department of Nuclear Science and Engineering, Massachusetts Institute of Technology, Cambridge, Massachusetts 02139, USA

[2] Department of Materials Science and Engineering, Massachusetts Institute of Technology, Cambridge, Massachusetts 02139, USA

## Abstract

Quantum anomalous Hall (QAH) effect generates quantized electric charge Hall conductance without external magnetic field. It requires both nontrivial band topology and time-reversal symmetry (TRS) breaking. In most cases, one could break the TRS of time-reversal invariant topological materials to yield QAH effect, which is essentially a topological phase transition. Conventional topological phase transition induced by external field/stimulus needs a route along which the bandgap closes and re-opens. Hence, the phase transition occurs only when the magnitude of field/stimulus is larger than a critical value. In this work we propose that using gapless surface states, the transition can happen at arbitrarily weak (but finite) external field strength. This can be regarded as an unconventional topological phase transition, where the bandgap closing is guaranteed by bulk-edge correspondence and symmetries, while the bandgap reopening is induced by external fields. We demonstrate this concept on the 2D surface states of 3D topological insulators like $Bi_2Se_3$, which become 2D QAH insulators once a circularly polarized light is turned on, according to van Vleck's effective Hamiltonian in Floquet time crystal theory. The sign of quantized Chern number can be controlled via the chirality of the light. This provides a convenient and dynamical approach to trigger topological phase transitions and create QAH insulators.

---

[†] correspondence to: liju@mit.edu

Light has become a powerful tool for tuning material behaviors without direct contact. A promising application explored in recent years is light-driven phase transitions. Compared with conventional mechanical, thermal, electrical, or electrochemical approaches to triggering phase transitions, using light as an external driving force is advantageous as light can be non-contact, non-destructive, and ultrafast. There are already several theoretical proposals and experimental observations that light (or focused laser pulses) can trigger structural phase transitions in both three-dimensional bulk materials and low-dimensional nanostructures[1–3]. In addition, light can also change the electronic structure and induce electronic phase transitions. In insulators or semiconductors, above-bandgap light can excite electron-hole pairs, and the system would acquire metallic properties in carrier transports. In a sense, this can be regarded as an insulator-to-metal transition. Indeed, the reverse process, metal-to-insulator phase transition under light, which is more counter-intuitive, has also been proposed[4].

In recent years, band topological orders (characterized by e.g., $Z_2$ number) has become an important paradigm for material classifications. To date, hundreds of materials have been predicted to possess nontrivial electronic band topologies[5–7], and some of them have already been fabricated and demonstrated in experiments. However, the quantum anomalous Hall (QAH) effect[8,9] characterized by the Chern number $\mathcal{C}$ ($\in \mathbb{Z}$), is still challenging for experimental observations. The QAH effect features an integer quantum Hall conductance $\sigma_{xy} = \mathcal{C}\frac{e^2}{h}$ without external magnetic field, where $e$ is the electron charge and $h$ is the Planck constant. Notably, QAH insulators are rare in nature, due to two stringent, necessary (but not sufficient) conditions, namely inverted band structures near the Fermi level and broken time-reversal symmetry (TRS)[9]. The first practical model for QAH insulators was proposed by Haldane[8]. Then it was demonstrated that introducing magnetic dopant atoms into topological insulators (TIs) could break the TRS and lead to the QAH effect[10]. In 2009, Chang et al. performed the first successful experiment and observed the QAH effect in Cr or V doped (Bi, Sb)$_2$Te$_3$ thin films at very low temperature (~80 mK)[11]. However, such extrinsic doping requires careful control over the impurity magnetic sites and interactions. Recently, it was found that intrinsic QAH effect can be observed in MnBi$_2$Te$_4$ thin films at an elevated temperature of 4 K[12–14]. Despite these advances, more experimentally accessible materials and novel mechanisms to realize and observe QAH effect at high temperatures need to be explored. In this work, we demonstrate quantum phase transition and 2D QAH effect on the gapless surfaces of 3D topological insulators like $Bi_2Se_3$ can be driven and controlled under light irradiation. We theoretically and computationally analyzed how the surface states evolve with light illumination and demonstrate how the anomalous Hall conductivity arises on the surfaces of TIs. We clarify that the Hall conductance under CPL only exists on the top and bottom layers (several quintuple layers) of the $Bi_2Se_3$ slab, while the middle layers remain silent. Our work could provide more detailed evidence for careful experimental verifications and potential applications. In addition, we propose an unconventional pathway for topological phase transitions. We point out that in principle, an arbitrarily small external field would be able to induce topological phase transitions in gapless systems, in contrast to the conventional topological phase transitions, where a finite and usually large external field is required. This unconventional topological phase transition is applicable in many gapless systems beyond the surface states of TIs.

Considering the interaction of electrons with monochromatic light of frequency $\Omega$, the electronic system has a time-periodic Hamiltonian $H(t) = H(t + T)$, where $T \equiv \frac{2\pi}{\Omega}$ is the period. The temporal periodicity is reminiscent of the spatial periodicity in crystals (translation), and can be systematically treated with the Floquet time-crystal theory[15–18] analogous to Bloch's theorem. Intuitively, there can be virtual *interactions* between the system at time $t$ and its temporal images at $t + mT$ ($m \in \mathbb{Z}$), similar to the interaction between an atom and its spatial image in neighboring unit cells. Such interaction provides a dynamical tool for tuning the properties of the system. When the periodic perturbation is weak, and its frequency $\Omega$ is much higher than the observational frequency (energy) scale of interest so that no resonant transitions can happen, one can apply the high-frequency (van Vleck's) expansion, and obtain an effective time-independent Floquet Hamiltonian,

$$H^{\text{F}} \approx \widetilde{H}^0 + \sum_{m \neq 0} \frac{[\widetilde{H}^{-m}, \widetilde{H}^m]}{2m\Omega} \tag{1}$$

where $\widetilde{H}^m = \frac{1}{T}\int_0^T dt\, H(t) e^{im\Omega t}$ is the Fourier transform of $H(t)$. Here we keep only the lowest-order terms in the Floquet expansion. Utilizing the Floquet theory, it has been demonstrated that the electronic structures of the materials can be controlled with light[19,20], and particularly, topologically trivial materials could become topologically nontrivial under light illumination without structural (ionic) changes[21–24]. For example, the anomalous Hall effect under CPL in graphene has been proposed[21] and observed recently[25]. It has also been proposed that light could induce effective spin-orbit coupling and trigger the quantum spin Hall to QAH transition in checkerboard antiferromagnetic superconductor FeSe monolayer[26]. However, the transition requires ultra-strong light with AC electric field strength on the order of 1 V/Å. Besides, free-standing FeSe monolayers are notoriously hard to fabricate. Hence, it is highly desirable to explore the light-induced QAH effect 1) under lower light intensity and 2) in materials with better experimental feasibility. In this work, we propose that under circularly polarized light (CPL) the 2D surface states of 3D $Z_2$-TIs could show QAH effect. A unique advantage of starting from the 2D surface states of TIs is that the intensity of the CPL required to trigger the quantum phase transition can be arbitrarily weak. This is because the 2D TI surface states are gapless by themselves and could easily transit to QAH insulators once their bandgaps are opened. This property may make experimental observations significantly easier. From a practical point of view, this could also reduce light absorption and the possible heating effects, especially when the light frequency is below the bulk bandgap and the electron-hole pair generation can be significantly reduced. Besides, the surface states are particularly sensitive to light at low frequencies (e.g., infrared or terahertz), and may find applications in light detection.

It is well known that topological electronic phase transitions can be triggered by external stimuli (denoted as $F$ here), such as strain, electric field, light, etc. Except for some rare cases[27], a common and prominent feature of conventional topological phase transitions is that the bandgap of the material needs to close at a critical strength of the external field ($F = F_{\text{cri}}$) and then reopen as the field strength further increases ($F > F_{\text{cri}}$). On two sides of such critical strength ($F < F_{\text{cri}}$ and $F > F_{\text{cri}}$), the system usually has different topological properties, and the topological phase transition occurs at $F = F_{\text{cri}}$ (Figure 1a). This picture holds in the case of the transition from time-reversal symmetric TIs to QAH insulators as well. The TRS breaking field (induced by magnetic doping, etc.) needs to reach a critical value $F_{\text{cri}}$ to trigger QAH phase. When the TRS breaking field is weak ($F < F_{\text{cri}}$), the Hall conductance remains zero in the system, even if the TRS is broken[28]. This is verified in the case of both magnetic doping[10] and CPL irradiation[26], where a finite critical external field strength is required to close and reopen the bandgap.

Notably, the magnitude of the critical field strength $F_{\text{cri}}$ is usually not small. For example, it is on the order of a few percent elastic strain[29], a field strength of 1 V/Å in the case of static electric field[30], and an intensity of $10^{12}$ W/cm$^2$ in the case of light-induced phase transitions[26]. Such large critical field strengths hinder the observation and applications of topological phase transitions, and may also induce unwanted impurities or effects. As discussed previously, when the field strength goes from $F = 0$ to $F = F_{cri}$, the bandgap reduces from the intrinsic value down to zero. Hence, a natural speculation is that, if the system is initially gapless, then the topological phase transition can happen at zero field strength ($F_{cri} = 0$), as the bandgap closing process is not required any more. In this case, once the external field is turned on (arbitrarily small strength), the system would immediately open its bandgap and transit to a different topological phase (Figure 1b). This can be understood with a thermodynamic phase transition picture. For the conventional field induced topological transitions shown in Figure 1a, the initial system is located at a local minimum on the transition path, thus a finite $F_{cri}$ is required to bring the system onto the (electronic) transition saddle point. On the other hand, if the topological phase transition starts from a gapless phase (Figure 1b), then the system is initially located on the transition saddle point, thus $F_{\text{cri}}$ can be arbitrarily small.

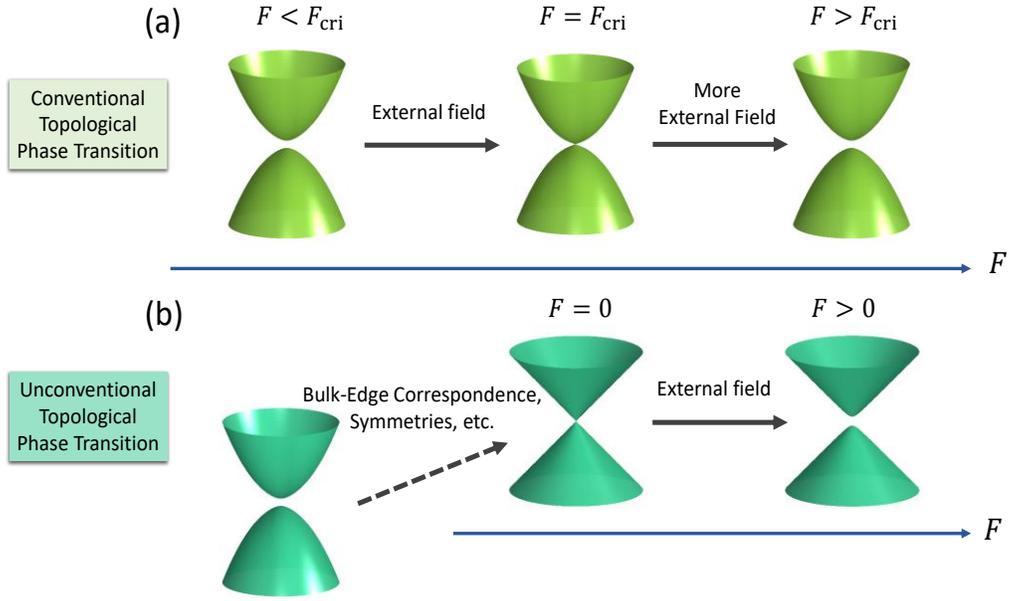

**Figure 1** Two types of topological phase transitions. (a) Conventional topological phase transition. The bulk bandgap closing and reopening processes are both triggered by external field $F$. And the topological phase transition occurs at a critical field strength $F_{cri}$, where the bandgap just closes. (b) Unconventional topological phase transition. The bandgap closing is guaranteed by bulk-edge correspondence, symmetries, etc., whereas the bandgap reopening is induced by external field. Once the bandgap is opened, the system can change into a different topological state.

Fortunately, some systems are guaranteed to be gapless. For example, the 2D surface states of a 3D $Z_2$-TI is protected to be gapless when interfaced with topologically trivial systems (such as vacuum). This is because when continuously connecting two systems with different band topologies, the bandgap must close in between. Such gapless surface states are robust against perturbations, disorders, and impurities that preserve TRS[31–33]. However, when TRS is broken, the surface bandgaps could open and the QAH effect may arise. Therefore, one could start from these gapless surface states and trigger the QAH effect with CPL, which breaks TRS. Note that this topologically protected gapless state is different from graphene, since the latter one is not immune to external doping and requires high-quality fabrication process. Here we first study this effect with a minimal model Hamiltonian[34] that can describe the surface states of TIs, $H_{\text{SS}}(\boldsymbol{k}) = \hbar v_{\text{F}}(k_x \sigma_y - k_y \sigma_x)$, for 2D Dirac Fermion. Here $v_{\text{F}}$ is the velocity of the Dirac Fermion, and $\sigma_i$ ($i = x, y, z$) are the Pauli spin matrices. One can derive the effective Floquet Hamiltonian (see Supplementary Materials, SM) under CPL irradiation as

$$H_{\text{SS}}^{\text{F}}(\boldsymbol{k}) = \hbar v_{\text{F}}(k_x \sigma_y - k_y \sigma_x) \pm \frac{e^2 v_{\text{F}}^2 A^2}{\hbar \Omega} \sigma_z \qquad (2)$$

where $+$ and $-$ correspond to right- and left-handed CPL, respectively. $\boldsymbol{A}$ and $\Omega$ are the vector potential and the frequency of the CPL, respectively. The last term is induced by the CPL and represents an exchange field that breaks the TRS and opens a bandgap of $\frac{e^2 v_{\text{F}}^2 A^2}{\hbar \Omega}$. It is well-known that Eq. (2) describes QAH insulators[8,35] with a Chern number of $C = \pm 1$. Note that here $\boldsymbol{A}$ can be arbitrarily small. Besides, these results remain the same when the higher order terms are incorporated in the $k \cdot p$ model to reflect the warping from real lattice symmetries[36].

To illustrate the above toy model in a real material, we take bulk Bi$_2$Se$_3$ as an example, which is a well-studied $Z_2$-TI[34,37]. First-principles density functional theory calculation is performed to reproduce its electronic band structure more accurately than the toy model Hamiltonian. The atomic structure of Bi$_2$Se$_3$ (Figure 2a) has a space group of $R\bar{3}m$ and has a layered structure along the $z$ direction. Each layer is constituted by five atom layers (Se-Bi-Se-Bi-Se), and is dubbed a quintuple layer (QL). The bulk of Bi$_2$Se$_3$ has a bandgap of ~0.3 eV [34,37],

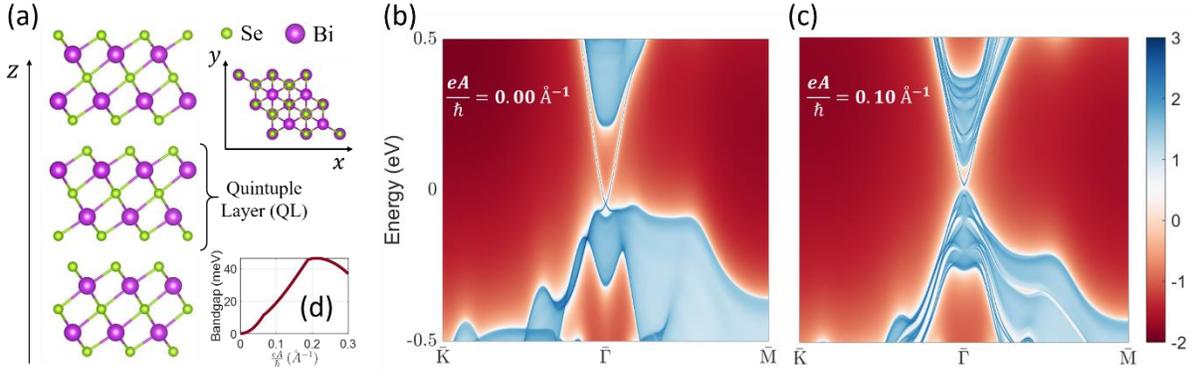

**Figure 2** (a) Atomic structure of $Bi_2Se_3$. (b, c) The surface state spectrum of $Bi_2Se_3$ under (b) no light and (c) CPL with field strength $\frac{eA}{\hbar} = 0.1$ Å$^{-1}$, respectively. A bandgap of 20 meV can be seen in (c). The colormap in (b, c) is in logarithmic scale. (d) Bandgap of the surface states of $Bi_2Se_3$ as a function of the CPL strength. The bandgaps are calculated with a slab model with 24 QLs.

while the surface states of $Bi_2Se_3$ are gapless. In Figure 2b we plot the electronic dispersion of its [111] surface, where a gapless Dirac dispersion at the $\bar{\Gamma}$ point can be seen. We now add a CPL propagating along the $z$ direction, with time-dependent vector potential $\boldsymbol{A}(t) = A(\cos\Omega t, \eta \sin\Omega t, 0)$, where $\eta = +1$ and $-1$ corresponds left- and right-handed CPL, respectively. We take $\hbar\Omega = 5$ eV for the following calculations, which is much higher than the frequency (energy) range of interest in this work. Here we would like to remark that a smaller frequency (especially below the bandgap ~0.3 eV) may be more favorable in experiments. We adopt $\Omega = 5$ eV mainly from a computational point of view, as the theoretical error from the van Vleck's expansion would be smaller at this high frequency. In Figure 2c we plot the surface spectrum function under left-handed CPL with intensity of $\frac{eA}{\hbar} = 0.1$ Å$^{-1}$ (corresponding to an electric field strength of $E = 5$ V/nm), which is calculated based on the Floquet formalism (see SM). A bandgap of $E_g \approx 20$ meV can be clearly observed. We then adjust the light intensity and explore its relationship with the bandgap. When the light is not too strong, the bandgap scales as $E_g \propto A^2 \propto I$, where $I$ is the light intensity (Figure 2d). This relationship is intuitive as $I$ characterizes the strength of TRS breaking. Also, the bandgap opening is a second-order nonlinear effect induced by the photo-dressing of the electronic states, hence $E_g$ should be linearly proportional to $I$, which is also proportional to the number of photons irradiated. However, for a strong light with $\frac{eA}{\hbar} \gtrsim 0.2$ Å$^{-1}$, the fundamental bandgap tends to decrease as the light intensity increases. When the light is strong enough, the system becomes metallic. This is due to the interplay between different orbitals in $Bi_2Se_3$, and is absent when one uses a low-energy effective Hamiltonian as described above, which involves only a subset of the atomic orbitals near the Fermi level. On the other hand, when a linearly polarized light is applied, which does not break TRS, the surface states remain gapless (SM Figure S1).

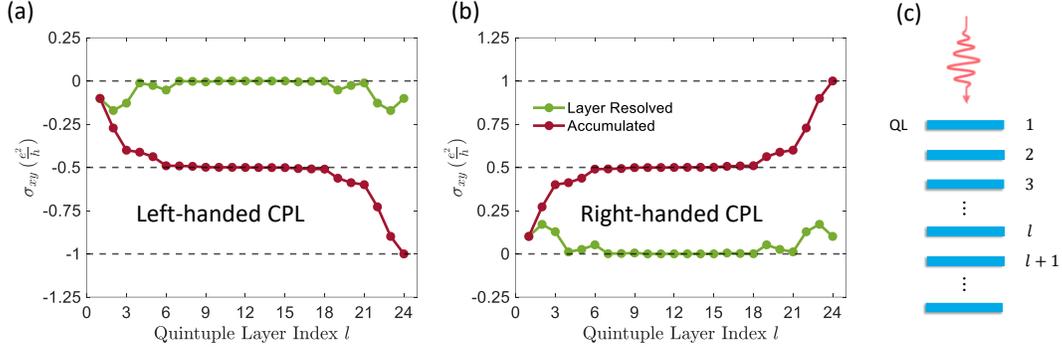

**Figure 3** Layer-resolved Hall conductance of a $Bi_2Se_3$ slab with 24 QLs under (a) left-handed and (b) right-handed CPL with $\frac{eA}{\hbar} = 0.1$ Å$^{-1}$. The top and bottom surface each contributes $0.5\frac{e^2}{h}$, and the whole slab is a QAH insulator. The green dots represent layer-resolved Hall conductance of each layer index by $l$, while the red dots are summation of Hall conductance from the 1$^{st}$ (top) to the $l^{th}$ layers. (c) A schematic illustration of the $Bi_2Se_3$ slab under light.

Usually, bandgap closing and opening correspond to topological phase transitions. To quantify the topological nature of the $Bi_2Se_3$ surface state under CPL, we calculate its Hall conductance, according to the Kubo formula,

$$\sigma_{ab} = \frac{e^2}{\hbar} \sum_{n \neq m} \frac{d\mathbf{k}}{(2\pi)^2} (f_n - f_m) \frac{\text{Im}\{\langle m|v_a|n\rangle\langle n|v_b|m\rangle\}}{(\omega_m - \omega_n)^2} \qquad (3)$$

Here the dependences of quantities on $\mathbf{k}$ are omitted in the notation. $|n\rangle$, $\omega_n$ and $f_n$ are the eigenstates, eigenvalues (band frequency), and occupancy of the $n$-th band of the Floquet Hamiltonian $H^F$, respectively. $v_a = \frac{1}{\hbar}\frac{\partial H^F}{\partial k_a}$ with $a = x, y$ is the velocity operator. A slab model is used to calculate the Hall conductance. Specifically, we first build an intrinsic Hamiltonian $H$ from *ab initio* calculations, and then evaluate the effective Hamiltonian $H^F$ under light using the Floquet formalism. In the current case, we need to resolve the contributions from different QLs in the system. We define a spatial projection operator $P_l = \sum_{i \in l} |\psi_i\rangle\langle\psi_i|$. Here $|\psi_i\rangle$ are atomic orbitals, and the summation runs over all orbitals centered on the $l$-th QL. Then we replace the current operator $v_a$ with $P_l v_a$, which corresponds to the current on the $l$-th QL. In this way, a layer-resolved conductance $\sigma_{ab}^l$ can be obtained. Note that by summing over $l$, the total conductance of the whole slab can be recovered. Here we would like to note that in the periodically driven system, the electrons are usually out-of-equilibrium. As a result, Eq. (3) should be considered as an approximation to the dynamical Hall conductivity, with a correction up to the order of $\mathcal{O}(A^2)$. Such an approximation requires[38] that 1) the frequency of the light is off-resonance so that no direct interband transitions occur; 2) the intensity of light $A^2$ is small. Both of them are satisfied in the present work. In general cases where resonant interband transitions can happen, or $A^2$ is large, the occupation of the Floquet bands can significantly deviate from the Fermi-Dirac distribution and one should apply the Floquet theory in a more rigorous fashion[39–41]. Generally speaking, the nonzero Hall conductivity is expected, but it may deviate from the quantized value, depending on the actual experimental conditions. Here we would like to note again that the choice of light frequency $\Omega = 5$ eV is mainly from a theoretical and computational standpoint, but the essence of our results holds true at lower frequencies. At low frequencies, a bandgap of 20 meV can be obtained at a lower electric field strength (see SM). Actually, in Ref. [42] a CPL with $\Omega = 0.12$ eV and $E = 2.5 \times 10^7$ V/m was used, and a bandgap of $E_g \approx 50$ meV was observed on the surfaces of Bi$_2$Se$_3$. Under these conditions, our *ab initio* calculation predicts a bandgap of $E_g \approx 35$ meV according to the van Vleck's expansion. This demonstrates that the van Vleck's expansion can give a qualitatively correct result even when the light frequency is small.

We first use a slab model with 24 QLs, which is thick enough to rule out the interaction between the top and bottom surfaces. Under left-handed CPL with $\frac{eA}{\hbar} = 0.1$ Å$^{-1}$, the calculated layer-resolved Hall conductance $\frac{1}{2}(\sigma_{xy}^l - \sigma_{yx}^l)$ is shown in Figure 3. Here the green dots represent the layer-resolved conductance, whereas the red dots show the total conductance measured from the first QL to the $l$-th QL, i.e., $\sum_{i=1}^{l} \frac{1}{2}(\sigma_{xy}^i - \sigma_{yx}^i)$. From Figure 3a, one can see that the whole slab system has a quantized Hall conductance of $-\frac{e^2}{h}$. From the layer-resolved Hall conductance plot, we find that only the top and bottom surfaces (roughly 6 QLs) contribute to $\sigma_{xy}$; each gives $-0.5 \frac{e^2}{h}$. Under right-handed CPL, the Hall conductivity flips its sign (Figure 3b). Thus, the system has Chern number $\mathcal{C} = +1$ and $-1$ under right- and left-handed CPL, respectively. This demonstrates that the whole slab becomes a QAH system when the bandgap is opened under CPL irradiation. Hence, the light serves as an effective SOC interaction. The strength and sign of this effective SOC can be fine-tuned via light intensity and handedness. This is different from the usual atomic SOC interaction, which is determined mainly by the atomic number and is also positive, leaving little room for tunability. On the other hand, QLs in the middle of the slab remain silent and have almost zero Hall conductance. We also calculated the spin Hall conductance, and find that each middle QL gives a spin Hall conductance of $0.36 \frac{\hbar}{2e} \frac{e^2}{h}$ (see SM), which is close to the layer-resolved spin Hall conductance in a bulk Bi$_2$Se$_3$. This again suggests that although the CPL breaks the TRS in the middle QLs, it is not strong enough to trigger the phase transition to the QAH insulator with the conventional pathway depicted in Figure 1a. However, the transition can be triggered on the surfaces with the unconventional pathway in Figure 1b, because one does not need a finite critical field to close the bandgap on the surfaces.

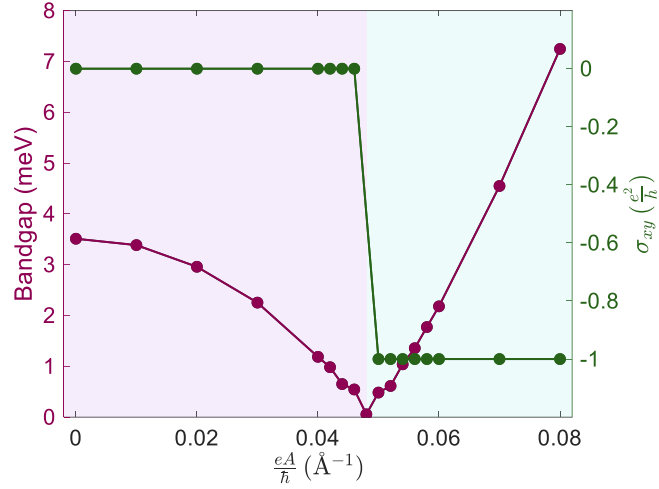

**Figure 4** The bandgap (left $y$-axis) and Hall conductance (right $y$-axis) of a Bi$_2$Se$_3$ slab with 3 QLs as a function of the field strength of the left-handed CPL. The system has finite bandgap and zero Hall conductance before light illumination. A transition to QAH insulator happens at $\frac{eA}{\hbar} \approx 0.07$ Å$^{-1}$, where the bandgap closes and reopens.

To better understand the thickness effect and explore the benefit of using the surface states, we consider a thin slab of Bi$_2$Se$_3$ as a comparison. In the thin-slab scenario, the quantum tunneling between the top and bottom surface states leads to a mass term and opens a bandgap. Specifically, the bandgap of a slab with 6 QLs is around 4 meV. When we turn on the CPL, the bandgap gradually decreases (Figure 4), but the Hall conductance remains zero until the bandgap closes at a critical field strength $\frac{eA_{\text{cri}}}{\hbar} \approx 0.05$ Å$^{-1}$. With $\frac{eA}{\hbar} > 0.05$ Å$^{-1}$, the system becomes a QAH insulator with Hall conductance of $-\frac{e^2}{h}$. This is a vivid illustration of the conventional topological phase transitions depicted in Figure 1a, which cannot happen below a critical field strength. When the thickness of the slab $\to \infty$, the critical field strength $\to 0$, and we recover our main proposition. On the other hand, if we use a thinner slab, then the critical light intensity required to trigger the phase transition would be even higher. For example, if we use 3 QLs, then the intrinsic bandgap is around 50 meV, and the transition to a QAH state cannot happen even when $\frac{eA}{\hbar}$ is 0.15 Å$^{-1}$.

Before concluding, we would like to make several remarks. First, the influence of CPL can also be interpreted as being caused by the inverse Faraday effect[43]. It is well-known that the CPL can induce an effective magnetic field (or equivalently, an effective magnetization), which would naturally induce a Hall conductance. This is also consistent with the analysis above that CPL induces an exchange interaction [Eq. (2)]. On the other hand, in Ref. [43] it was demonstrated that not only a CPL but also a linearly polarized light could lead to a nonzero static magnetization. This occurs when the frequency of the linearly polarized light is above the electronic bandgap. In this case, the electron interband transitions would cause energy dissipations, which breaks the TRS according to the second law of thermodynamics[43]. From this point of view, a linearly polarized light might also lead to the QAH effect on the surfaces of topological materials, provided that the dissipations are taken into consideration[43,44]. Such possibility will be studied in future work.

Second, the unconventional topological phase transition may be used for light detection, especially in the low-frequency range (e.g., terahertz). As discussed above, the bandgap opened by the CPL is $E_g \propto \frac{A^2}{\Omega} \propto \frac{I}{\Omega^3}$, where $I$ and $\Omega$ are intensity and frequency of the light, respectively. The $E_g \propto \Omega^{-3}$ scaling law indicates that the surface states are particularly sensitive to light with relatively low frequencies. This should be compared with conventional approaches for light detection, whose sensitivity decreases as the light frequency decreases. Low frequencies below the bulk bandgap have another advantage that the absorption in the bulk can be avoided. Of course, at very low frequencies, one should use the Floquet theory in a more formal way than the high-frequency expansion used in the current work, but the stronger sensitivity at low frequencies is qualitatively true. In addition, at very low frequencies, other unwanted processes, such as the coupling with phonons, may come into play. A thorough consideration of these effects will be the focus of future work. Besides, the sharp jump of the Hall conductance $\sigma_{xy}$ from 0 to $\frac{e^2}{h}$ can be detected by optical approaches such as magneto-optical Kerr or Faraday rotation, which can make a possible all-optical light detection device.

Finally, this unconventional topological phase transition may also apply to other gapless systems, such as the surface states of topological crystalline insulators[45,46] or Dirac semimetals[47]. In these systems, the zero bandgaps are protected by crystal spatial symmetries, hence the topological properties may strongly couple with phonons, which can break certain crystal symmetries. Thus, these systems could be ideal platforms for studying phonon-dressed electronic states dynamically.

In conclusion, we demonstrate that the 2D surface states of 3D TIs can transit to QAH insulators under circularly polarized light (CPL). A prominent feature is that the critical light strength required to trigger the surface electronic-state phase transition can be arbitrarily small since the surface bandgap is already closed according to the 3D bulk-edge correspondence. Such an unconventional topological phase transition can make easier experimental observations of QAH effects under high temperature and may find practical applications such as light detection. Intuitively, one may think of the CPL as generating effective magnetism on the gapless surface, thus giving rise to the QAH effect.

**Acknowledgments**

This work was supported by the Office of Naval Research Multidisciplinary University Research Initiative Award No. ONR N00014-18-1-2497.